\documentclass[10pt,final,journal]{IEEEtran}

\IEEEoverridecommandlockouts

\usepackage{cite}
\usepackage{amsmath,amssymb,amsfonts,newtxmath}
\usepackage{graphicx,xcolor,booktabs}
\usepackage[caption=false,font=footnotesize]{subfig}

\newcommand{\expval}[1]{\operatorname{E}[#1]}

\begin{document}
	
\title{Performance Prediction for Coherent Noise Radars Using the Correlation Coefficient}

\author{
	\IEEEauthorblockN{David Luong,~\IEEEmembership{Graduate Student Member, IEEE}, Bhashyam Balaji,~\IEEEmembership{Senior Member, IEEE}}, and \\ Sreeraman Rajan,~\IEEEmembership{Senior Member, IEEE}
	
	\thanks{D.\ Luong is with Carleton University, Ottawa, ON, Canada K1S 5B6. Email: david.luong3@carleton.ca.}
	\thanks{B.\ Balaji is with Defence Research and Development Canada, Ottawa, ON, Canada K2K 2Y7. Email: bhashyam.balaji@drdc-rddc.gc.ca.}
	\thanks{S.\ Rajan is with Carleton University, Ottawa, ON, Canada K1S 5B6. Email: sreeramanr@sce.carleton.ca.}
}

\maketitle

\begin{abstract}
	Noise radars can be understood in terms of a correlation coefficient which characterizes their detection performance. Although most results in the literature are stated in terms of the signal-to-noise ratio (SNR), we show that it is possible to carry out performance prediction in terms of the correlation coefficient. To this end, we derive the range dependence of the correlation coefficient. We then combine our result with a previously-derived expression for the receiver operating characteristic (ROC) curve of a coherent noise radar, showing that we can obtain ROC curves for varying ranges. A comparison with corresponding results for a conventional radar employing coherent integration shows that our results are sensible. The aim of our work is to show that the correlation coefficient is a viable adjunct to SNR in understanding radar performance.
\end{abstract}


\section{Introduction}

Noise radar, as the name suggests, uses a noise waveform as its transmit signal \cite{thayaparan2006noise,kulpa2013signal,narayanan2016noise,wasserzier2019noise,cooper1967random}. As in other radars, noise radars retain a copy of the transmitted signal as a reference for matched filtering. Due to the presence of noise, the reference signal is not necessarily a perfect copy of the signal that was transmitted through free space. Relatively little attention has been paid to the degradation of the reference signal used for matched filtering. Often, the degradation is assumed to be arbitrarily small, as was done in \cite{dawood2001roc} for example.
	
This motivates the use of the Pearson correlation coefficient between the free-space signal and the reference signal, or the \emph{correlation coefficient} for short. Although this correlation has appeared in previous publications on noise radar \cite{cooper1967random,milne1993theoretical,dawood2001roc}, most results in noise radar have been stated in terms of the signal-to-noise ratio (SNR) of the free-space signal. This is because SNR is far more familiar to engineers. However, the correlation coefficient takes into account both the degradation of the free-space signal and the degradation of the reference signal, the latter of which is not captured by the SNR \cite{dawood2001roc,luong2020snrdet}. We feel, therefore, that the correlation coefficient is a sensible metric for evaluating the performance of noise radars---and perhaps of other radars, too. Moreover, the correlation coefficient is an easy way to highlight the surprising connection between noise radar and the new field of quantum radar \cite{luong2019roc,luong2020magazine,luong2019cov}.

In this paper, we show how performance prediction for a coherent noise radar can be carried out in terms of the correlation coefficient. We use the radar range equation to derive the dependence of the correlation coefficient on the range of a given target; this equation holds irrespective of whether the radar is coherent or not. We then exploit a previous result, which relates the receiver operating characteristic (ROC) curve of a coherent noise radar to the correlation coefficient, to show how the ROC curve for a noise radar varies with range.

\section{The Noise Radar Protocol}
\label{sec:summary}

In our analysis, we will consider radars which work as follows:
\begin{enumerate}
	\item Produce two correlated zero-mean Gaussian random noise signals.
	\item Retain one of the signals and send it directly to the receiver for use as a reference for matched filtering. Transmit the other signal toward a target.
	\item Measure the in-phase and quadrature voltages at the receiver.
	\item Correlate the received and recorded signals. Declare a detection if the correlation exceeds a given threshold.
\end{enumerate}
Normally, the reference signal in step 2 would be digitized immediately upon generation if it were not generated digitally in the first place. In this case, the received signal in step 3 would also be digitized and the correlation would be performed via digital signal processing. An example of the practical implementation of such a scheme can be seen in \cite{stove2016design}. It is interesting to note, however, that the reference signal could be sent to the receiver in analog form and the correlation in step 4 performed via analog signal processing, as done in \cite{dawood2001roc}. Our theoretical results would not change in either case, but for simplicity we will assume that we are working with digital signal processing.

\section{The Noise Radar Correlation Coefficient}
\label{sec:corr_coeff}

Let us denote the time series of in-phase and quadrature voltages of the received signal by $I_1(t)$ and $Q_1(t)$, respectively. Similarly, let us denote the voltages of the reference signal by $I_2(t)$ and $Q_2(t)$. We model these voltage time series as stationary, zero-mean, real-valued Gaussian white noise processes that are mutually uncorrelated when the time difference between the signals is nonzero. Therefore, we will drop the time variable for simplicity and assume that the time difference is always zero. We will assume further that the target is stationary, the phase shift between transmit and receive is a constant (which implies that the radar is coherent), and that any system or external noise is additive white Gaussian noise.

Under these assumptions, the four voltages are completely characterized by the covariance matrix $\expval{xx^T}$ where $x = [I_1, Q_1, I_2, Q_2]^T$. It was shown in \cite{dawood2001roc} (though in different notation) that, if the reference signal is a direct copy of the transmitted signal, the covariance matrix can be written in block matrix form as
\begin{equation} \label{eq:cov_mat}
	R_\text{QTMS}(P_1, P_2, \rho, \phi) = 
	\begin{bmatrix}
		P_1 \mathbf{1}_2 & \rho \sqrt{P_1 P_2} \mathbf{R}(\phi) \\
		\rho \sqrt{P_1 P_2} \mathbf{R}(\phi) & P_2 \mathbf{1}_2
	\end{bmatrix}
\end{equation}
where $P_1$ and $P_2$ are the powers (or, equivalently, the variances) of the received and reference signals respectively, $\mathbf{1}_2$ is the $2 \times 2$ identity matrix, $\rho$ is a parameter such that $0 \leq \rho \leq 1$, $\phi$ is the phase, and $\mathbf{R}(\phi)$ is the rotation matrix
\begin{equation}
	\mathbf{R}(\phi) = 
	\begin{bmatrix}
		\cos \phi & \sin \phi \\
		-\sin \phi & \cos \phi
	\end{bmatrix} \! .
\end{equation}
Alternatively, the transmitted and reference signals can be generated by mixing a single source of bandlimited Gaussian noise with a carrier signal. This results in two sidebands of correlated noise, one of which can be transmitted and the other retained. This case was considered in \cite{luong2019cov}, where it was shown that the form of the resulting correlation matrix is the same as \eqref{eq:cov_mat} except that instead of a rotation matrix, a reflection matrix appears instead:
\begin{equation}
	\mathbf{R}'(\phi) = 
	\begin{bmatrix}
		\cos \phi & \sin \phi \\
		\sin \phi & -\cos \phi
	\end{bmatrix} \! .
\end{equation}
It is noteworthy that, when $\mathbf{R}'(\phi)$ is used in \eqref{eq:cov_mat}, it is not sensible to form the complex voltages $z_1 = I_1 + j Q_1$ and $z_2 = I_2 + j Q_2$ as is standard in many engineering applications. This is because $\expval{z_1 z_2^*} = 0$, so conventional matched filtering would have no effect. This is why we prefer to work with real-valued voltages.

\subsection{Target Detection and the Correlation Coefficient}
\label{subsec:detection}

The parameter $\rho$ in \eqref{eq:cov_mat} is the focus of this paper. We call it the \emph{correlation coefficient} because it characterizes the strength of the correlation between the received and transmitted signals. This can be seen by noting that, when the phase shift $\phi$ is zero, $\expval{I_1 I_2} = \rho \sqrt{P_1 P_2}$. In this case, $\rho$ is simply the Pearson correlation coefficient between $I_1$ and $I_2$. The effect of $\phi$ is to ``distribute'' the correlation among the cross-covariances $\expval{I_1 I_2}$, $\expval{I_1 Q_2}$, $\expval{Q_1 I_2}$, and $\expval{Q_1 Q_2}$. Note that we can always choose $\rho \geq 0$ because its sign can be absorbed into $\mathbf{R}(\phi)$ or $\mathbf{R}'(\phi)$.

The correlation coefficient is strongly related to the problem of target detection. At one extreme, if $\rho = 1$, then the received and reference signals are perfectly correlated. This would occur in the ideal case where there is absolutely no noise introduced into the signal. On the other hand, if $\rho = 0$, then the two signals are completely uncorrelated. This would be the case if there were no target at all, so the received signal is not an echo of the transmitted signal.

The above discussion suggests that detecting a target with a noise radar reduces to distinguishing between the following two hypotheses:
\begin{alignat*}{3}
	H_0&: \rho = 0 &&\quad\text{Target absent} \\
	H_1&: \rho > 0 &&\quad\text{Target present}
\end{alignat*}
Therefore, we can think of $\rho$ as a detector function. In terms of the correlation coefficient, step 4 of the protocol described in Sec.\ \ref{sec:summary} can be more concretely stated as follows: calculate an estimate $\hat{\rho}$ of the correlation coefficient of the radar's received and recorded signals, set a threshold, and declare a detection if $\hat{\rho}$ lies above the threshold. The method we use for calculating the correlation coefficient is to perform the minimization
\begin{equation} \label{eq:minimization}
	\min_{P_1, P_2, \rho, \phi} \left\lVert R_\text{QTMS}(P_1, P_2, \rho, \phi) - \hat{S} \right\rVert_F
\end{equation}
subject to the constraints $0 \leq P_1$, $0 \leq P_2$, $0 \leq \rho \leq 1$, and $0 \leq \phi \leq 2\pi$. Here $\hat{S}$ is the sample covariance matrix calculated directly from the radar's voltage measurements and $F$ denotes the Frobenius norm.  The value of $\rho$ that minimizes this expression is taken as the estimate $\hat{\rho}$. Full details of this procedure, together with approximations for ROC curves as a function of the given parameters, may be found in \cite{luong2019rice}. Note that this minimization is where the coherence of the radar comes into play: if the radar were incoherent, minimizing \eqref{eq:minimization} over $\phi$ would not be meaningful and the resulting $\hat{\rho}$ may not be accurate.

\section{Correlation Coefficient as a Function of Range}

Given the importance of the correlation coefficient, its range dependence is of considerable interest. We therefore present a theoretical derivation here. In order to simplify the analysis, it is convenient to assume that the received and reference signals can be decomposed into perfectly correlated and perfectly uncorrelated parts. That is, we assume that each signal is the sum of a component which is common to both signals and a component which is independent of the other signal. The uncorrelated parts can be thought of as the total amount of noise in each signal. We can then write
\begin{subequations}
\begin{align}
	P_1 &= P + P_{n1} \\
	P_2 &= P + P_{n2}
\end{align}
\end{subequations}
where $P$ is the power of the perfectly correlated part while $P_{n1}$ and $P_{n2}$ are the powers of the uncorrelated parts of the two signals. Implicit in this decomposition is the assumption that both signals have the same power at the source. This need not be true, but this does not affect our analysis because any gain factor in one or another of the signals would cancel out in the following calculations. We also note that the above decomposition is a mathematical abstraction, as the signals may be tainted with noise at the very source and there may exist no perfectly correlated \emph{physical} signal. However, this decomposition is useful because it was shown in \cite{luong2019cov} that the correlation coefficient $\rho$ can be written in the form
\begin{equation} \label{eq:rho}
	\rho = \left[ \left( 1 + \frac{P_{n1}}{P} \right) \! \left( 1 + \frac{P_{n2}}{P} \right) \right]^{\! -\frac{1}{2}} \! .
\end{equation}
A similar expression is derived in \cite{dawood2001roc} except that their results were in terms of SNR, whereas we aim to develop a theory for the correlation coefficient without resorting to SNR.

In order to obtain the range dependence of $\rho$, we first eliminate $P$ in favor of $P_1$ and $P_2$ to give
\begin{align} \label{eq:rho_rewritten}
	\rho &= \left[ \left( 1 + \frac{P_{n1}}{P_1 - P_{n1}} \right) \!\! \left( 1 + \frac{P_{n2}}{P_2 - P_{n2}} \right) \right]^{\! -\frac{1}{2}} \nonumber \\
		&= \sqrt{\left( 1 - \frac{P_{n1}}{P_1} \right) \! \left( 1 - \frac{P_{n2}}{P_2} \right)}
\end{align}
Next, we assume that $P_{n1} > P_{n2}$. This is reasonable because the received signal will be contaminated with both system noise and external noise, whereas the reference signal is contaminated only with system noise. We also assume that, if there were no external noise, the two signals would be contaminated with the same amount of noise. Therefore we can write
\begin{equation}
	P_{n1} = P_{n2} + P_n,
\end{equation}
where $P_n$ is the power of the external noise added to the received signal. It follows from \eqref{eq:rho_rewritten} that, if there were no external noise, the correlation coefficient would be
\begin{equation} \label{eq:rho0}
	\rho_0 = 1 - \frac{P_{n2}}{P_2}.
\end{equation}
This represents the maximum correlation that can be observed by the radar. The fact that it is less than unity is a reflection of the fact that the radar contains system noise.

According to one form of the radar range equation \cite{skolnik1962introduction}, the received power $P_1$ can be written as
\begin{equation} \label{eq:power1}
	P_1 = \frac{G A_e \sigma}{(4\pi)^2 R^4} P_2 + P_{n},
\end{equation}
where $G$ is the gain of the transmit antenna, $A_e$ is the effective area of the receive antenna, $\sigma$ is the target's radar cross section (RCS), and $R$ is the range. (If another form of the radar range equation is desired, it can be used here in an entirely analogous way.) We have used the assumption that both the reference and transmitted signals have the same power at the source, as noted earlier. The power of the noise component within the received signal may similarly be written as
\begin{equation} \label{eq:noisepower1}
	P_{n1} = \frac{G A_e \sigma}{(4\pi)^2 R^4} P_{n2} + P_n.
\end{equation}
Finally, substituting equations \eqref{eq:rho0}, \eqref{eq:power1}, and \eqref{eq:noisepower1} into \eqref{eq:rho_rewritten} yields 
\begin{equation} \label{eq:rho_range}
	\rho(R) = \frac{\rho_0}{\sqrt{1 + (R/R_c)^4}}
\end{equation}
where $R_c$ is a characteristic length defined as
\begin{equation}
	R_c = \left( \frac{G A_e \sigma P_2}{(4\pi)^2 P_n} \right)^{\! 1/4}.
\end{equation}
It is the range at which the received signal power is equal to the received noise---in other words, the range at which SNR is unity (0~dB). For pulsed radars that do not rely on matched filtering, $R_c$ would be the theoretical maximum range. It is also the range at which the correlation coefficient is reduced to $1/\sqrt{2}$ of its maximum value. For the example parameters given in Table \ref{table:parameters}, which were inspired by the values given in \cite{stove2016design}, we find that $R_c = 1.0$~km.

\begin{table}
	\caption{Example parameters}
	\label{table:parameters}
	\centerline{
		\begin{tabular}{ccc}  
			\toprule
			Parameter & Variable & Value \\
			\midrule
			Tx antenna gain & $G$ & 30 dB \\
			Rx antenna effective area & $A_e$ & 0.081 m\textsuperscript{2} \\
			Target RCS & $\sigma$ & 1.0 m\textsuperscript{2} \\
			Tx signal power & $P_2$ & 18 dBm \\
			Rx noise power & $P_n$ & $-$94 dBm \\
			\bottomrule
		\end{tabular}
	}
\end{table}

\begin{figure}[t]
	\centerline{\includegraphics[width=\columnwidth]{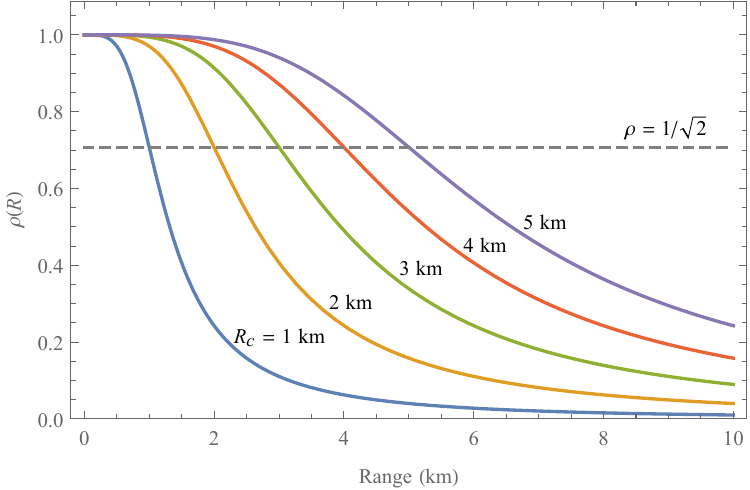}}
	\caption{Correlation coefficient as a function of range for varying values of the characteristic length $R_c$, assuming $\rho_0 = 1$. Dashed line indicates $\rho = 1/\sqrt{2}$.}
	\label{fig:rho_range}
\end{figure}

We have thus obtained the range dependence on the correlation coefficient in terms of two quantities which are relatively easy to determine: the initial correlation $\rho_0$ and the characteristic length $R_c$. The former is a measure of the best possible performance that the radar can deliver, and does not depend on anything outside the radar. It can be thought of as the quality of the matched filtering performed by the radar. The latter is essentially the radar range equation, and can be thought of as a measure of how quickly the performance of the radar decays with range. Because none of the derivations in this section depend on the phase between the transmitter and receiver, \eqref{eq:rho_range} holds for all noise radars, whether coherent or not.

Fig.\ \ref{fig:rho_range} plots the correlation coefficient as a function of range for varying values of $R_c$. Since $\rho_0$ appears in \eqref{eq:rho_range} as a multiplicative constant, our plots show only the case where $\rho_0 = 1$.

\section{Receiver Operating Characteristic Curve}

As mentioned in Sec.\ \ref{subsec:detection}, $\rho$ can be considered a detector function for use in the problem of distinguishing between the presence or absence of a target. In this section, we study the ROC curves that are obtained when $\rho$ is used as the detector function.

The procedure we use to estimate $\rho$ from the voltage time series $I_1$, $Q_1$, $I_2$, and $Q_2$ was briefly described in Sec.\ \ref{subsec:detection}. For that procedure, there exists an explicit expression for the ROC curve \cite{luong2019rice}:
\begin{equation} \label{eq:ROC}
	p_\text{D}(p_\text{FA} | \rho, N) = Q_1 \! \left( \frac{\rho \sqrt{2N}}{1 - \rho^2}, \frac{\sqrt{-2 \ln p_\text{FA}}}{1 - \rho^2} \right) \! .
\end{equation}\textbf{}
Here $p_\text{D}$ is the probability of detection, $p_\text{FA}$ is the probability of false alarm, $N$ is the number of voltage samples over which to integrate, and $Q_1$ denotes the Marcum $Q$-function (not to be confused with the quadrature voltage $Q_1$ of the signal received by the radar). This is an approximate expression that holds when $N$ is greater than approximately 100.

\begin{figure}[t]
	\centerline{\includegraphics[width=\columnwidth]{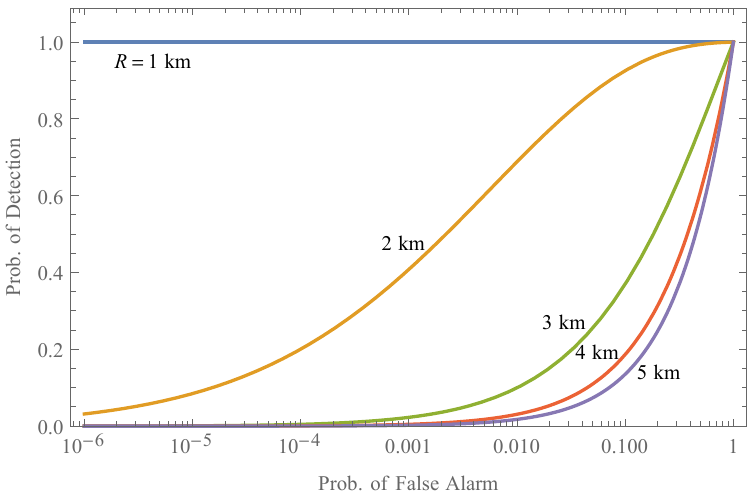}}
	\caption{ROC curves for varying ranges, assuming $\rho_0 = 0.8$, $R_c = 1.0$~km, and $N = 150$.}
	\label{fig:ROC_range}
\end{figure}

The range dependence of the ROC curve is obtained simply by substituting \eqref{eq:rho_range} into \eqref{eq:ROC}. Representative plots are shown in Fig.\ \ref{fig:ROC_range}. {We take $R_c = 1$~km because that is the result obtained from the parameters in Table \ref{table:parameters}. As might be expected, the probability of detection falls precipitously as $R$ becomes significantly larger than $R_c$.

We note that there are various other detector functions which could be used, such as the envelope detector studied in \cite{dawood2001roc} as well as the one described in \cite{luong2020simdet}. The ROC curves for both of these also depend on the correlation coefficient, so it is possible to determine detector performance as a function of range for both detectors by substituting \eqref{eq:rho_range}. In fact, the ROC curve expressions in \cite{dawood2001roc} have the merit of being valid for all $N$, not just for large $N$ as described above. We do not analyze it here because the expressions are difficult to work with even numerically.

As a check on the plausibility of our expressions, we compare the ROC curves obtained here with those for a conventional coherent radar using a sinusoidal waveform as described in, e.g., \cite{mahafza2000radar}. Recalling that $R_c$ is the range at which $\mathit{SNR} = 1$ and that power varies inversely with the fourth power of the range, it follows that the single-pulse SNR is
\begin{equation}
	\mathit{SNR} = \left( \frac{R_c}{R} \right)^{\! 4}.
\end{equation}
It is known that the ROC curve for such a conventional radar, assuming perfect coherent integration, is given by
\begin{equation} \label{eq:ROC_conventional}
	p_\text{D}(p_\text{FA} | \mathit{SNR}, N) = Q_1 \Big( \! \sqrt{2 N \cdot \mathit{SNR}}, \sqrt{-2 \ln p_\text{FA}} \Big) .
\end{equation}
A derivation of \eqref{eq:ROC_conventional} can be found in \cite{mahafza2000radar}. Curiously, the Marcum $Q$-function appears both in \eqref{eq:ROC_conventional} and in \eqref{eq:ROC}, though we repeat that the latter is an approximate result which holds only for large $N$, whereas \eqref{eq:ROC_conventional} is exact. Plots of ROC curves for various SNRs, for both noise radar and conventional radar, are given in Fig.\ \ref{fig:ROC_conventional}. Note that, in the derivation of \eqref{eq:ROC_conventional}, noise is assumed to be added only to the received signal \cite{mahafza2000radar}. We therefore take $\rho_0 = 1$ in our comparison of conventional radar with noise radar, so the noise radar performs perfect matched filtering and there is no noise in the reference signal.

It can be seen that, for the most part, the ROC curves are comparable between the two radars. The differences between the two arise from the different waveforms employed by the two radars and the use of $\rho$ as a detector function for the noise radar, which is not analogous to the envelope detector of a conventional radar. Because of these differences, we would expect only a rough correspondence between the ROC curves of the two types of radars. Nevertheless, the two cases are similar enough to show that the results we have obtained are reasonable.

\begin{figure}[t]
	\centerline{\includegraphics[width=\columnwidth]{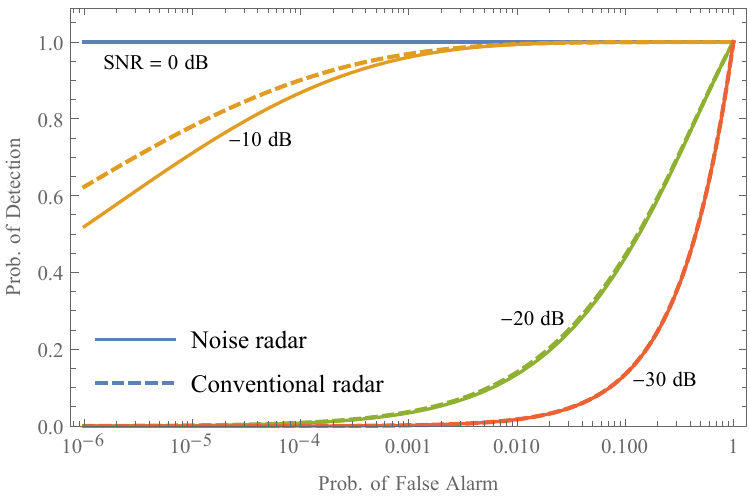}}
	\caption{ROC curves for various SNRs, assuming $\rho_0 = 1$ and $N = 150$.}
	\label{fig:ROC_conventional}
\end{figure}

\section{Conclusion}

In this paper, we saw that noise radars can be described in terms of a certain correlation coefficient $\rho$ which is intimately related to detection performance. We then derived the range dependence of this coefficient. This result holds whether or not the radar is coherent. Finally, we showed that when the radar is coherent, we can obtain ROC curves for varying target ranges by combining the range dependence with a previously derived expression for the ROC curve.

Much exploratory work remains to be done to show how changes in $\rho$ would affect radar performance. For example, it would be of interest to determine the Cram\'{e}r–Rao bound for bearing estimation in terms of $\rho$. Another important question is the exact relationship between $\rho$ and SNR. We also aim to calculate the values of $\rho$ and $N$ required to achieve desired values of $p_\text{D}$ and $p_\text{FA}$, in a similar fashion to what was done in \cite{dawood2001roc}. Other directions for future work include generalizing our results to cases where the assumptions listed in Sec.\ \ref{sec:corr_coeff} do not hold, such as moving targets, time-varying phase shifts between transmit and receive, non-Gaussian additive noise, and multiplicative noise scenarios.

We also plan to explore the applicability of noise radar to various sensing applications. For example, biomedical sensors may benefit from the fact that noise waveforms are less likely to interfere with other medical equipment compared to the sinusoidal waveforms used in many radars. In particular, we could explore the applicability of noise radar to fall detection \cite{sadreazami2020fall}. The performance prediction framework presented above could help us to understand the detection performance we could expect from a fall detector based on noise waveforms. Our work could also help us decide whether the enhanced detection performance of quantum radars would be helpful for fall detection or other sensing applications.

We hope that in our paper, we have been able to show that the correlation coefficient is a viable lens through which the performance of noise radars can be understood.


\bibliographystyle{ieeetran}
\bibliography{qradar_refs,own_refs}
	
\end{document}